\documentclass[letterpaper]{article}
\usepackage{arxiv}

\usepackage[utf8]{inputenc} 
\usepackage[T1]{fontenc}    
\usepackage{multirow}
\usepackage{cite}
\usepackage{hhline}
\usepackage{graphicx}
\usepackage{subfigure}
\usepackage{setspace}
\doublespace
\usepackage{stfloats}
\usepackage{color}
\usepackage{cite}
\usepackage{amsmath, amsfonts, amssymb}
\usepackage{epstopdf}
\graphicspath{{./fig/}}

\begin{document}

\title{Blind Curvelet based Denoising of Seismic Surveys in Coherent and Incoherent Noise Environments}
\author{Naveed Iqbal,  Mohamed Deriche, Ghassan AlRegib, and Sikandar Khan
}
\maketitle

\begin{abstract}

The
localized nature of curvelet functions, together with their  frequency and dip  characteristics, makes the curvelet transform
an excellent choice for processing seismic data. In this work, a denoising method is proposed based on a combination of the curvelet transform and a whitening filter along with procedure for noise variance estimation. The whitening filter is added to get the best performance of the curvelet transform under coherent and incoherent correlated noise cases, and furthermore, it simplifies the  noise estimation method and makes it easy to use the standard threshold methodology without digging into the curvelet domain.  The proposed method is tested on pseudo-synthetic data by adding noise to  real noise-less data set of the Netherlands offshore F3 block and on the field data set from east Texas, USA, containing ground roll noise. Our experimental results show that the proposed algorithm can achieve the best results under all types of noises (incoherent or uncorrelated or random and coherent noise).
\end{abstract}
\vspace{-.35cm}
\section{Introduction}
Seismic surveys consist of large volumes of sensors. These sensors are laid on the earth surface to image the local geology for possible oil and gas reservoirs.  Sound waves are produced using a source (vibrator). The waves bounces off various layers within the earth and reflected waves to the surface are captured by sensors. The recorded data is usually  contaminated with unwanted energy or noise.  One of the main challenges in processing seismic data is to remove/suppress noise in order to enhance the Signal-to-Noise Ratio (SNR) and later, avoid the false interpretation of seismic data \cite{Shirzad2015,Diaz2015,Jones2014}, which can be detrimental.
 In general, seismic data processing can never eliminate noise completely, hence, the objective is to improve the SNR as much as possible.  Successful removal of the noise results in an  enhanced image of the subsurface geology, which facilitates
economical decisions such as  in hydrocarbon exploration.
In \cite{Ulrych1999}, authors define the signal of interest as the signal energy that is coherent from trace to trace and desirable for  interpretation from geophysical prospective. 
Noise includes disturbances in seismic data caused by any unwanted seismic energy, such as multiples, shot generation ground roll, effects of weather (wind, rain),  and human activities. Noise can also occur because of random occurrences within the earth. Broadly speaking, noise in seismic are classified as random noise,   coherent noise and correlated noise  \cite{Karsli2008,Aghayan2016}.

\begin{figure}[ht!]
\centering
  \includegraphics[width=8cm]{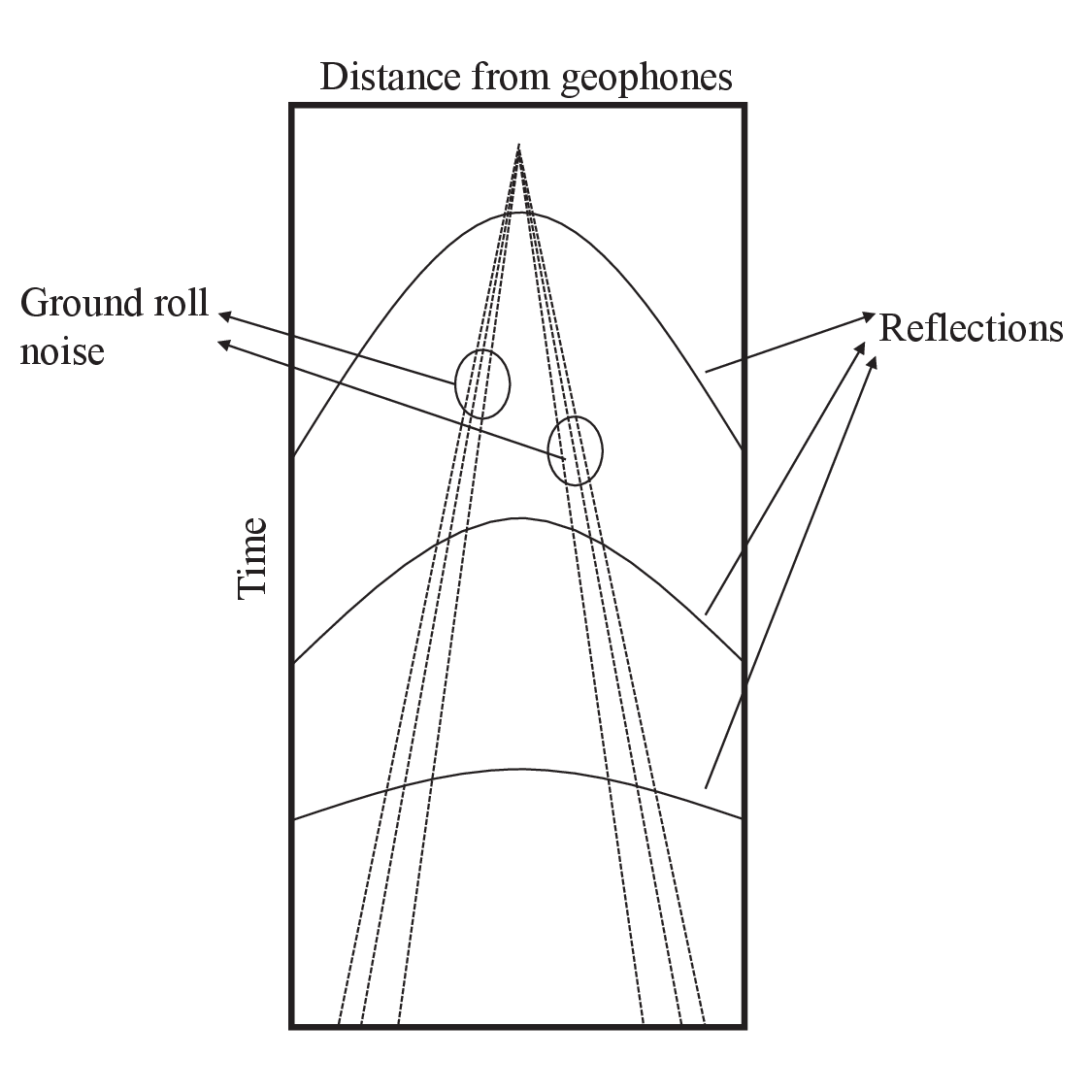}\\
  \caption{A pictorial view of ground roll noise and reflections.}
  \label{fig:intro}
\end{figure}

In short,  noise in seismic surveys can be classified as follows: \begin{itemize}
\item Incoherent noise: It is independent of the seismic source signal and there is no noise correlation among the traces.
\item Correlated and uncorrelated noise: Incoherent noise can have correlation or no correlation among the noise samples within the trace and, hence, called as correlated (Browian noise \cite{Bormann}, pink noise \cite{D.Peters2009} and noise through geophone \cite{HonsMichaelS.andStewart}) or uncorrelated (or random or additive white noise \cite{Duval2000}) noise, respectively. 
\item Coherent noise: It is generated by the source and it is highly correlated with the source signal and across the traces. Figure \ref{fig:intro} shows a case of coherent noise \cite{Rastegar2016}. This noise is the most troublesome, since it can be highly correlated  and aliased with the signal of interest\cite{Ozbek2000}.  
\end{itemize}

Whatever are the causes or types of seismic noise, these result in significant artifacts that negatively impact the subsequent  interpretation results, such as structural and spectral attributes, seismic imaging and analysis \cite{Kearey,Alcalde2017,Marfurt2015,Jedrzejowska-Tyczkowska2015}. 


In this work,  a novel and  robust method is proposed based on the curvelet-transform using a data-whitening step  which is able to deal with diverse and most challenging types of noise. The pre-whitening stage together with noise estimation from observation is collectively called as a blind curvelet based denoising method. The method is independent of type of noise and can be applied to variety of data sets. Therefore, the term `` blind" is used to indicate that noise type is not needed to be known and method can be applied blindly. In summary, the main contributions of this work are as follows:

\begin{itemize}
\item 
The performance of the curvelet denoising is classified based on the types of images. For this purpose, seismic images and general purpose images are compared. 
 \item A method to estimate noise from the seismic images for the coherent  and incoherent noise cases is presented based on the pattern of the images.
This simple and straightforward  method is in contrast with previous methods in which curvelets need to be identified and nullified to remove the noise. \item It is shown that curvelet transform gives best performance for the white noise (uncorrelated). For the colored (correlated) noise, there is still room for improvement. 
\item Inherently, the curvelet transform works well for uncorrelated noise. For performance enhancement, pre-whitening and whitening inverse steps are introduced for the correlated noise case. 
 \item Finally, the method which includes pre-whitening, noise estimation from the image and whitening inverse is adopted for the highly coherent case of ground roll noise.
\item Synthetic, pseudo-synthetic and field data sets are used to demonstrate the effectiveness of the proposed curvelet transform based denoising method.
\end{itemize}

\section{Related Work}
In this section, some of the  noise removal/reduction techniques for seismic data are briefly discussed.  

Band pass filtering and spectral filtering are  common techniques used for enhancing the SNR in seismic applications. However, if signal and noise share the same frequency content then either one  of these methods will result in signal attenuation and/or  noise will not be fully removed.
However, several other more complicated denoising methods have been proposed.   The author in \cite{Baziw2005}  discussed a particle-filter-based technique to enhance the SNR.  When compared with other signal filtering methods, the particle filter offers an advantage of broad-band signal cleansing by directly modeling the internal dynamics of the target physical system and statistical characteristics of the signal and noise. In doing so, signal filtering can be related to the dynamic characteristics of the underlying physical system, rather than a purely mathematical operation.
Deng \textit{et al.}\cite{Deng2015} discussed time-frequency peak filtering with an adaptive preprocessing to deal with the rapidly changing seismic wavelet. Furthermore, a threshold is designed to identify the noisy signal as a signal, buffer, or noise. In conventional time-frequency peak filtering, the Pseudo Wigner-Ville Distribution (PWVD) used for estimating the instantaneous frequency was shown to be sensitive to noise interferences that mask the borderline between the signal and noise, hence, detracting energy concentration.  The authors in \cite{Baziw2002} presented a real-time Kalman filter, which removes the statistically described background noise from the recorded seismic traces. However, statistics of noise are not available in reality. Cai \textit{et al.} \cite{Cai2011} used the generalized S-transform to transform seismic data from the time-space to the time-frequency-space (t-f-x) domain. Then, the Empirical Mode Decomposition (EMD) is applied on each frequency slice to suppress the coherent and random noise. An algorithm for wavelet denoising was proposed in \cite{Ghael1997} which used a wavelet-shrinkage estimate as a mean to design a wavelet-domain Wiener filter. The shrinkage estimate indirectly yields an estimate of the signal subspace that is leveraged into the design of the filter. The authors in \cite{Lu2006} proposed an adaptive Singular-Value Decomposition (SVD) filter to enhance the non-horizontal events by the detection of seismic image texture direction and then horizontal alignment of the estimated dip through data rotation. The features derived from the co-occurrence matrix are used to estimate the texture direction.
This technique works for random noise, however, some artifacts may appear.

Wiener filtering has been known for many decades and is considered to yield the optimum filter in the mean square sense. However, application of the Wiener filter directly for denoising problems in the seismic data requires statistical knowledge of the  seismic source signal or noise  which is not available in reality\cite{Coughlin2014}. A different approach is to assume a particular type of noise, e.g, white Gaussian noise   \cite{Aghayan2016} or model noise as a linear/non-linear sweep signal for ground-roll noise elimination \cite{Karsli2008}.

In this work, our aim is to develop a robust technique able to deal with diverse and most challenging types of noise, such as coherent noise (e.g. ground roll noise). It is worth mentioning that the denoising of ground roll noise is performed traditionally using band pass filtering \cite{Mousa2011a}. This increases the SNR\ by attenuating the ground roll noise, however, the seismic data is also effected by this frequency filtering. The reason being that the band that is excluded by the filtering process may contain important seismic events. Consequently, the vertical resolution of the seismic data is lowered. To alleviate this,
 Oliveira \textit{et al.}\cite{Oliveira2012} proposed a curvelet based denoising method using hard thresholding. In this method, the authors identify the angular sections that contain the ground roll noise and erase the corresponding curvelet coefficients before reconstructing the seismic signal. The problem with this technique is that it requires visual inspection of the angular sections to identify the  curvelet coefficients that are to be nullified to reduce noise. This process is demanding as it requires dedicated manpower to deal with the huge amount of data. A similar approach is also proposed in \cite{Neelamani2008}. Recently, G\'orszczyk \textit{et al.} \cite{Gorszczyk2014,Gorszczyk2015} proposed enhancement of 2-D and 3-D post-stack seismic data acquired in hardrock environment using 2-D curvelet transform. In their work, authors tackle random and colored noise. Scale-adjusted thresholding is introduced, which is a modification of \cite{Jean-LucStarck2002}, i.e., three levels are used for thresholding curvelet coefficients instead of two.  
\section{Curvelet Transform}
The transform-domain denoising methods are the most effective techniques 
in seismic  applications. These methods  achieve  the  goal of   noise removal or reduction  by  utilizing  the properties of  
sparseness and separateness present in seismic data in the transform domain. Wavelet transforms are popular in scientific and engineering fields, however, they ignore the geometric properties of structures, regularity of edges and curvature nature of edges. The curvelet transform\cite{Candes2006} is a multi-resolution directional transform which allows sparse representation of objects with curved edges. In comparison with the Fourier transform (sparse in frequency-domain and use translation only) and the wavelet transform (sparse in time and frequency-domains and use translations and dilation), the curvelet transform is sparse in time, frequency and phase-domains, and uses translations, dilation and rotation. 

In other words, the Fourier transform expresses a given time-domain signal, in terms of the amplitude and phase of each of the frequencies that are present in it. The wavelet transform is localized in both time and frequency whereas the standard Fourier transform is only localized in frequency. On the other hand, the curvelet transform is a higher dimensional generalization of the wavelet transform, which is  designed to represent images at different scales and different angles. It basically overcomes the problem of missing directional selectivity of  the wavelet transforms in 2-D signals (images).
Another transform, known as the Gabor transform, also uses different directions and scales like curvelet. However, curvelets completely cover the whole spectrum, whereas, there are many holes (blank region) in the frequency plan of the Gabor filters \cite{Kourav2013}.

There have been several other techniques of  directional wavelet systems
in recently proposed that have same goal, i.e.,  an optimal representation of directional features  and a better analysis of signals in higher dimensions.  These method include steerable wavelets, Gabor wavelets, wedgelets, beamlets, bandlets, contourlets, shearlets,
wave atoms, platelets, and surfacelets (these directional wavelets are uniformly called as X-lets). However, non of these methods has reached the same publicity as the curvelet transform. A good comparison of the above mentioned methods with curvelet can be found in    \cite{Ma2009}. In what follows, we will provide the mathematical background for the curvelet transform. The curvelet transform was initially developed by Cand\`es and Donoho \cite{Candes2000} and widely used for seismic data processing \cite{Long2018,Li2017,Li2017a,Gorszczyk2014,Gorszczyk2015}. The curvelets are represented by a triple index, i.e., scale $j$, spatial location $k=(k_1,k_2)$, and orientation $l$. A curvelet at any scale  $j$ can be thought of an oriented object that has a support in the rectangle having width and length of $2^{-j}$ and $2^{-j/2}$, respectively, and  obeys the parabolic scaling rotation width $\approx length^2$\cite{Candes2006}. The curvelet elements are obtained by parabolic translation, dilation and rotation of the specific function $\Phi_{j,l,k}$.
The mother curvelet $\Phi_j$ is defined in the frequency-domain as follows:
\begin{align}
\hat{\Phi}_j(r,\omega)=2^{-3/4}W(2^{-j}r)V\left( {2^{\lfloor j/2 \rfloor} \omega \over 2\Pi } \right),
\end{align}
where $(r,\omega)$ are the polar coordinates in the frequency-domain, $\lfloor j/2 \rfloor$ is the largest integer equal to or less than $j/2$, $W$ is the radial window and $V$ is the angular window. $W$ and $V$ are non-negative, smooth and real-valued functions and furthermore, these functions restrict the support $\hat{\Phi}_j$ to a polar wedge which is symmetric with respect to zero.

\begin{figure}
\centering
\includegraphics[width=9.3cm]{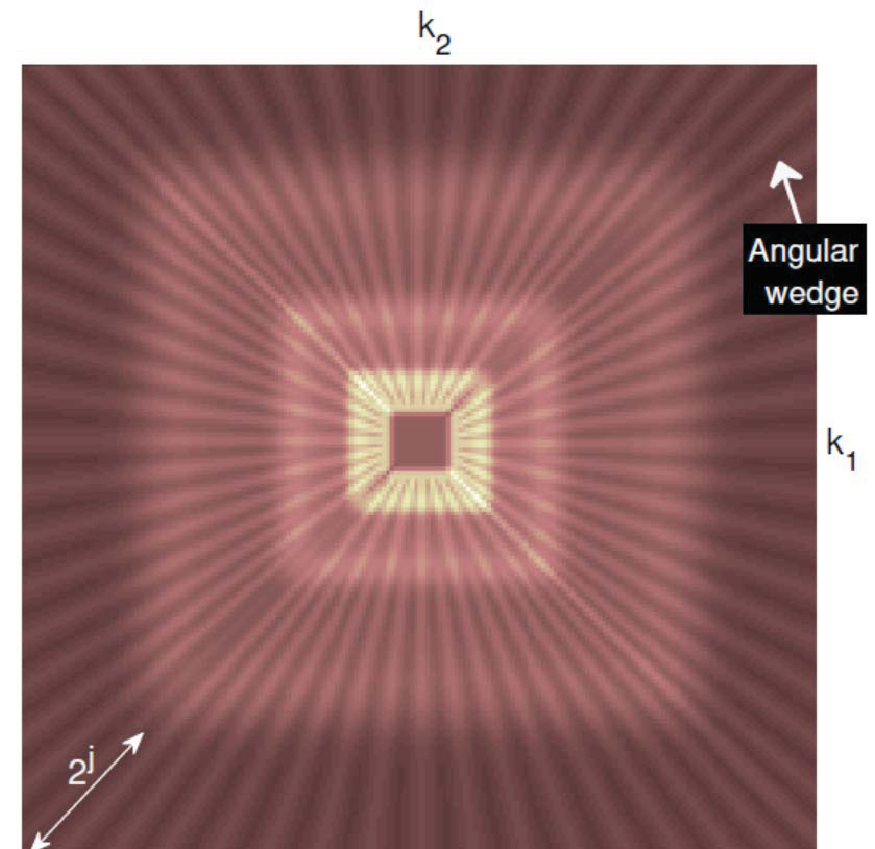}
\vspace{0.2cm}
  \caption{Default curvelet tiling.}
  \label{fig:curvelet}
\end{figure}

The curvelet ${\Phi}_{j,l,k}$ in the frequency-domain at scale $2^{-j}$, orientation $\psi_l$ and position $p_k^{j,l}$ is defined as:
\begin{align}
\hat{\Phi}_{j,l,k}=\hat{\Phi}_j(R_{\psi_l}\zeta)exp(i\langle p_k^{j,l},\zeta\rangle),
\end{align}
 where $p_k^{j.i}=R^{-1}_{\psi_l}(2^{-j}k_1,2^{-j/2}k_2)$ and $\zeta=(\zeta_1,\zeta_2)$ are the cartesian coordinates in the frequency-domain.
the rotation matrix $R_{\psi_l}$ is given as:
\begin{align}
R_{\psi_l} = \left(
\begin{array}{cc}
cos \ \psi_l & sin \ \psi_l \\
-sin \ \psi_l &cos \ \psi_l \\
 \end{array}
\right).\end{align}
The curvelet transform of a function $f$ is given by the convolutional integral:
\begin{align}
c_{j,l,k}=\int f(x)\overline{\Phi_{j,l,k}(x)} =\langle f,\Phi_{j,l,k} \rangle dx.
\end{align}
The coefficients of the curvelet transform, $c_{j,l,k}$, in the above equation are interpreted as the decomposition of function $f$ into the basis curvelet $\Phi_{j,l,k}$.
The curvelet-domain frequency-domain view is shown in Fig. \ref{fig:curvelet}.
 
The filtering methodology works as follows. First, the image or 2D signal is transformed into the curvelet-domain, then thresholding is applied and finally, signal is reconstructed after noise removal. For removing the noise from the image using the curvelet transform, we apply the standard methodology \cite{Jean-LucStarck2002,Neelamani2008,Nguyen2017,Shi2015}\cite{Oliveira2012} which is outlined here for the sake of self-containedness and clarity. Let assume that the noisy data is represented as
\begin{align}
x_{m,n}=I(m,n)+\sigma n_{m,n}
\end{align}
where $I$ is the image of interest and $n$ is the white noise with mean zero and variance $1$, i.e. $n_{m,n}\backsim N(0,1)$. Let $F$ denotes the discrete curvelet transform matrix, then, we have $Fn \backsim N(0,FF^T)$. Since the computation of the transform domain variance $FF^T$ is prohibitively
expensive, therefore, it is  calculated by an approximate value $\tilde \sigma_{\lambda}^2$ (where $\lambda$ is the curvelet index) of
the individual variances using Monte-Carlo simulations where
the diagonal elements of $FF^T$ are simply estimated by evaluating
the curvelet transforms of a few standard white noise images.
Let $c_\lambda$ be the  curvelet coefficients ($c=Fx$) corresponding to noise. The following hard-thresholding rule is used for removing noise
\begin{align}
\hat{c}_\lambda &= c_\lambda,\qquad \text{if} \qquad |c_\lambda|/\sigma \geq k\tilde \sigma \\
\hat{c}_\lambda & = 0,\qquad \ \text{if} \qquad |c_\lambda|/\sigma < k\tilde \sigma
\end{align} 
In our experiments, similar to \cite{Jean-LucStarck2002}, we actually chose a  value
for $k$ which is scale-dependent; hence, we have $k=4$ for the coarsest (first) scale $(j=1)$, whereas $k=3$ for
the other scales $(j>1)$.
\section{Correlated Noise and Whitening Method}
Curvelet transform based denoising performs most effectively under white noise scenario.
The reason being that the threshold assumes white noise and that the noise samples are uncorrelated. For colored (correlated) noise, the performance is not as good as for uncorrelated noise (this is shown in the results section). In \cite{Oliveira2012}, the authors discussed the case of correlated noise, e.g., ground roll noise, however, they intuitively find the curvelet coefficients to be eliminated. To the best of our knowledge, this issue of  threshold under the correlated noise case has not been addressed so far and  previous works mainly used  traditional methods.

Therefore, in this section, we introduce  a simple yet effective technique to deal with this issue. We propose to use a whitening procedure  to pre-whiten the data before curvelet-based processing in the case of colored noise. Here, we use the famous Zero-Phase Component Analysis (ZCA) for whitening. We start with Principal Component Analysis (PCA).  Suppose, $X$ is the image to be whiten. Note that, the columns of $X$ are correlated, hence, in case of coherent correlated noise columns represent  traces, whereas, in case of coherent noise rows represent traces. First, mean of each column is subtracted from the respective column. An estimate of the covariance matrix of the image $X$ is then obtained as:
\begin{align}
X^TX=U \Lambda U^T
\end{align}
where  $U$ is the square matrix whose $i^{th}$ column is the eigenvector $u_i$  of $X^TX$ and $\Lambda$ is the diagonal matrix whose diagonal elements are the corresponding eigenvalues. The whitening matrix $W_p$ then becomes
\begin{align}
W_{p} = U\Lambda^{-1/2}
\end{align}
and the whitened image is:
 \begin{align}
X_w=XW _{p}=XU\Lambda^{-1/2}
\end{align}
This transformation first rotates the variables using the eigen matrix 
$U$ of the covariance of $X$. This results in orthogonal components
but with, in general, different variances. To achieve whitened data the rotated
variables are then scaled by the square root of the eigenvalues $\Lambda^{1/2}$. PCA whitening
is probably the most widely applied whitening procedure due to its simplicity and relation to
PCA.
It can be seen that the PCA and ZCA whitening transformations are related by a
rotation matrix $U$, so ZCA whitening can be interpreted as a rotation followed by scaling followed
by the rotation $U$ back to the original coordinate system. Here note that the whitening by PCA is not unique. Any rotation (multiplying with a orthogonal rotation matrix) will leave it whitened. Hence, taking in particular $U$ from the covariance matrix and multiplying with the whitening matrix $W_p$ obtained from PCA, makes it unique. The new whitening matrix $W_z$ thus created form basis for ZCA. Therefore, the ZCA-whitening is given as

 \begin{align}
X_w=XW_{z} =XU\Lambda^{-1/2}U^T
\end{align}
$\Lambda^{-1/2}$ is a diagonal matrix with $1/(\sqrt \lambda_i)$ on the diagonal. This is regularized to $1/(\sqrt \lambda_i + \epsilon)$, where $ \epsilon)=0.0001$. The procedure for the case of correlated noise is as follows: 

\begin{itemize}
\item The image is whitened by multiplying it with the whitening matrix $W_z$.

\item Noise is estimated from the data (this will be discussed in the experimental results section)

\item Then, the whitened data  is denoised using the curvelet transform.

\item Finally, the whitened denoised data is restored by multiplying with the pseudo-inverse of the whitening matrix, i.e., $(W_z^HW_z)^{-1}W_z^H$.\end{itemize}
\section{Experimental results and Noise Estimation}
In this section, we will perform some experiments to test the proposed curvelet transform method and present our noise estimation approach based on the pattern of seismic images. 

In the first experiment, the curvelet denoising performance is compared for general purpose images and seismic images. In the  second experiment, the method to estimate noise from the seismic images for incoherent noise (correlated or uncorrelated) is introduced. In the third experiment, performance of curvelet transform is tested for correlated noise case, while  pre-whitening/whitening inverse steps are introduced in the fourth experiment in order to enhance the performance of curvelet based denoising under the correlated noise scenario. Finally, in the last experiment, worst case of ground roll noise is tackled using the whitening/whitening inverse filters and noise estimation from the image. 
\begin{figure}[ht!]
\centering
  \includegraphics[width=8.2cm]{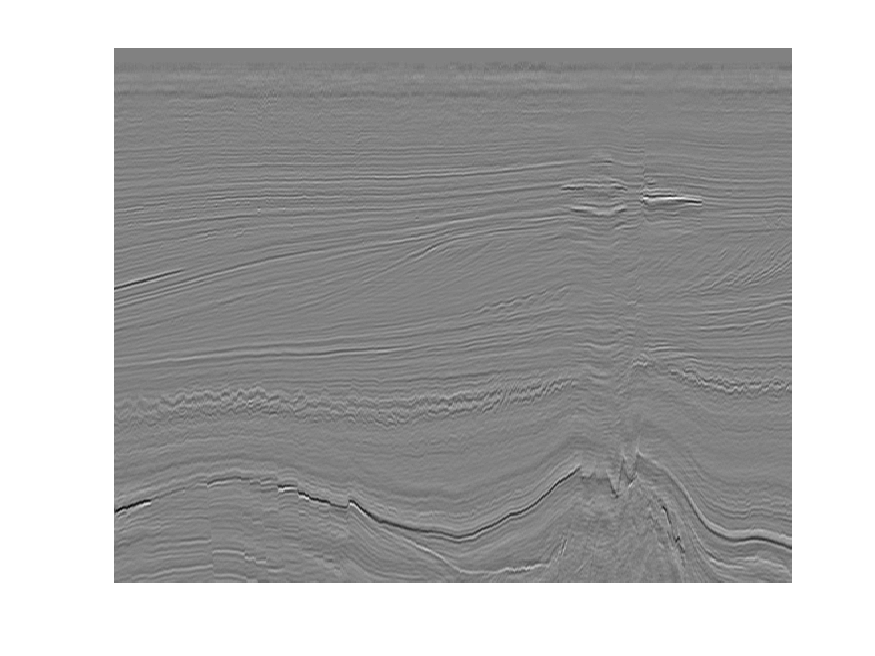}\\
  \caption{Samples of seismic images from the Netherlands
offshore F3 block  acquired in the North Sea with a resolution of $651 \times 951 \times 463$ (Crossline $\times$ Inline $\times$ Time).}
  \label{fig:s_im}
\end{figure}

\subsection{Performance of curvelet transform for general purpose images and seismic images}
\begin{figure}[ht!]
\centering
  \includegraphics[width=9cm]{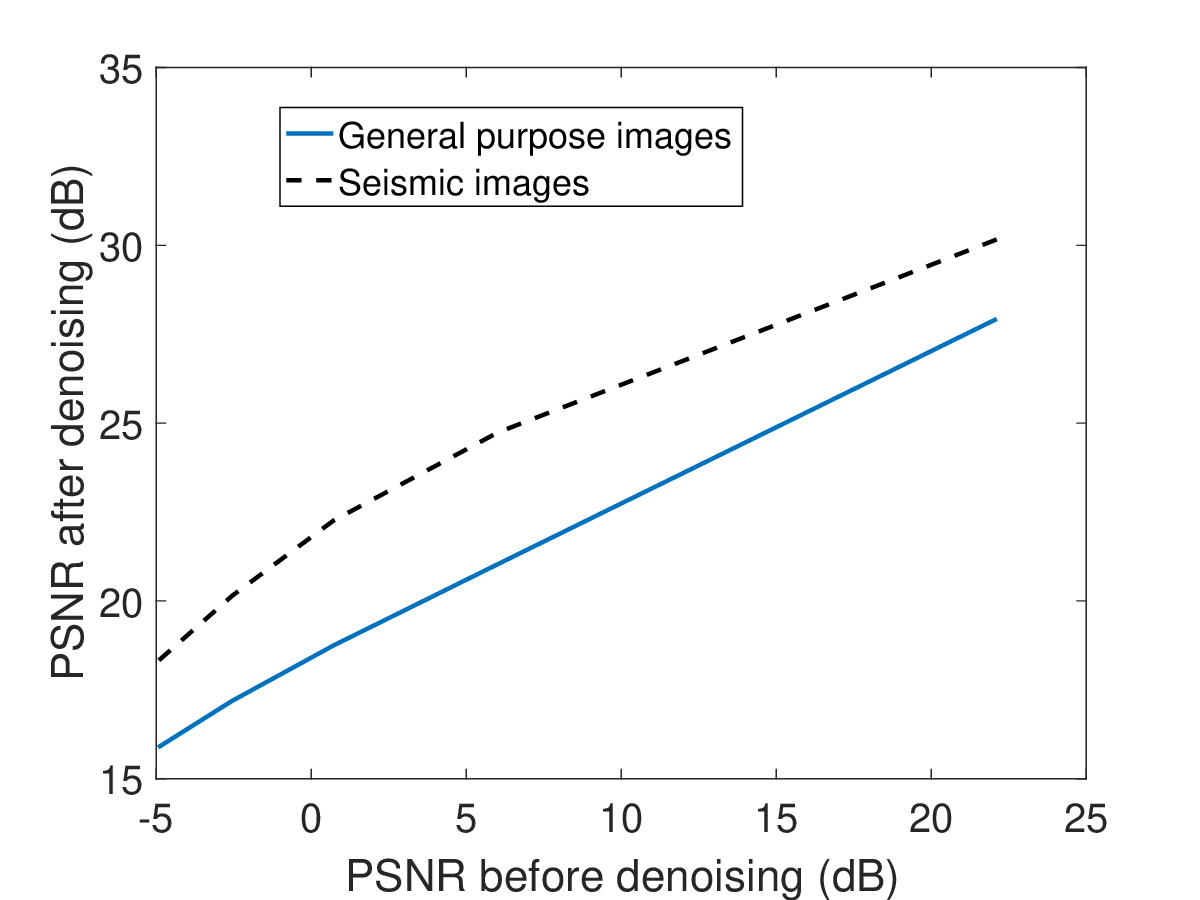}\\
  \caption{Performance of curvelet based denoising for general purpose images and seismic images.}
  \label{fig:psnr}
\end{figure}

In the first experiment, we compare the curvelet transform denoising performance for general purpose images and seismic images. An image of seismic section is shown in Fig.  \ref{fig:s_im}. For seismic images, we have used  the Netherlands
offshore F3 block that is acquired in the North Sea having a resolution of $651 \times 951 \times 463$ (Crossline $\times$ Inline $\times$ Time). From this data set, twenty images are taken from inline $\# \ 30$ to $600$ with a gap of $30$ inline sections.  White noise is added to this data set. The comparison of the two types of images is depicted in Fig. \ref{fig:psnr}, which reveals that the performance of curvelet transform based denoising is better for seismic images. The reason being that the seismic data are more localized in time, frequency, and phase domain and better match with the abilities of curvelet transform, i.e., taking into account the regularity  and curvature nature of edges. The Peak-Signal-to-Noise Ratio (PSNR) is calculated as in \cite{Al-Marzouqi2017}.
For general-purpose images, the authors assume that noise variance is known. 
\subsection{Noise variance estimation}
\begin{figure}
\centering
  \includegraphics[width=9cm]{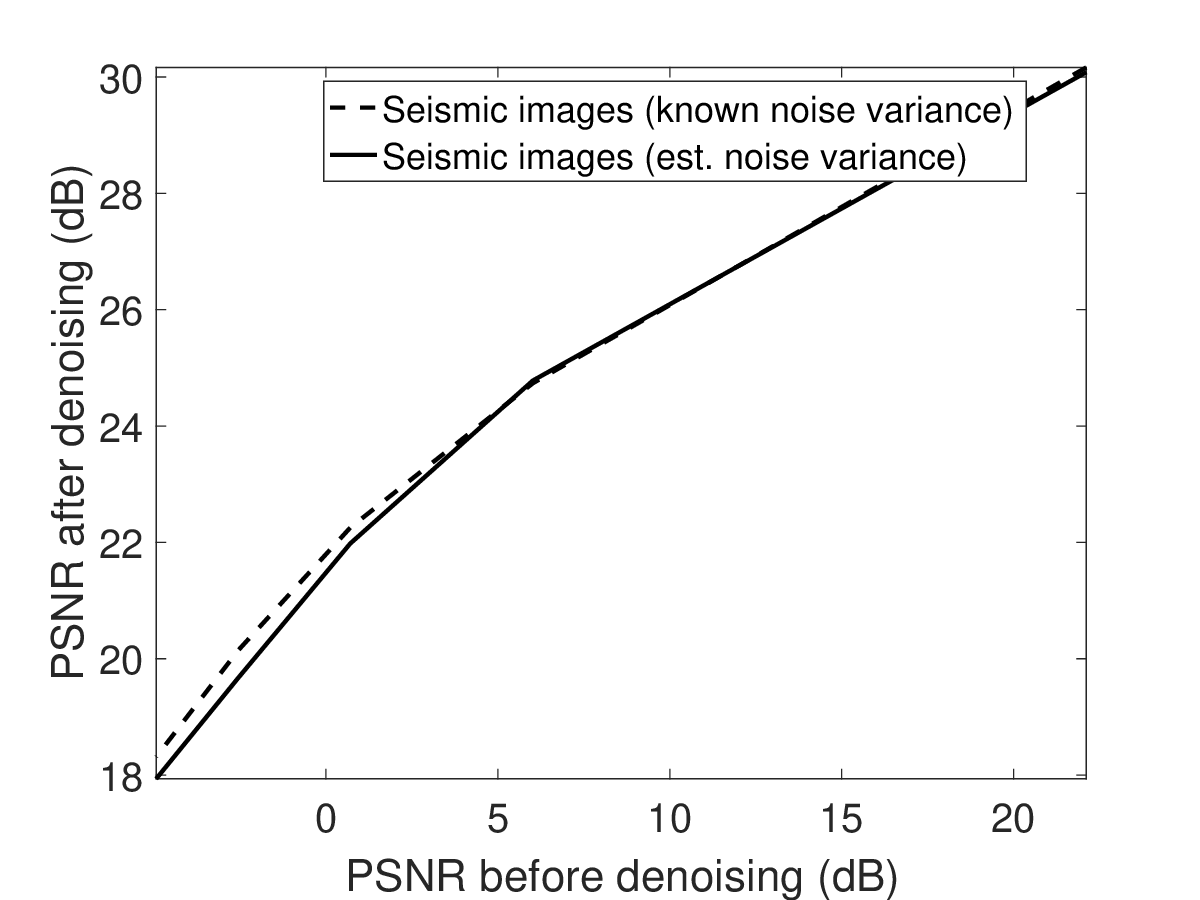}\\
  \caption{PSNR improvement for  seismic images with known and estimated noise variance.}
  \label{fig:psnr_est}
\end{figure}

The value of the noise variance is needed for thresholding in the curvelet domain \cite{Neelamani2008}. Previous works (mentioned before) about the curvelet-based denoising rely on the a priori known value of the noise variance. However, in reality the variance is not known and hence, curvelet transform-based denoising need to be performed using hit and try method. This, in turn, adds to the complexity of already heavy  computational load of the  curvelet transform. In  previous experiments, the variance was assumed to be known. Here, we estimate the variance using the image first $20 \times 20$ patch. Due to the pattern of the seismic image, the assumption here is  that there is no signal of interest (seismic signal) in this patch, which  is quiet true. The assumption fits in the case of noise that is not generated by the source signal. The comparison of the PSNR for known and estimated noise variance is shown in Fig. \ref{fig:psnr_est}.
The figure depicts that the curvelet based denoising achieves almost the same performance with known noise variance and the estimated one. These excellent results with noise variance estimation motivated us to use it for the rest of our experiments. 
\subsection{Correlated Noise}
\begin{figure}
\centering
  \includegraphics[width=9.5cm]{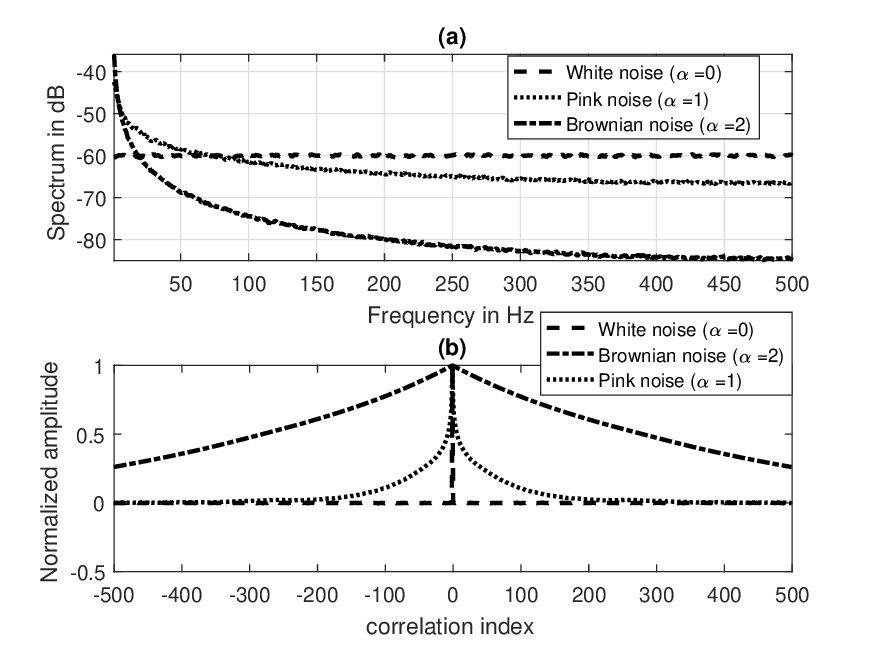}\\
  \caption{Noise specifications, (a) power spectral density, (b) autocorrelation function.}
  \label{fig:specs}
\end{figure}

In the aforementioned experiments white noise is used and curvelet-based denoising gives promising results. It is important to check its performance in correlated noise, which is likely to be present in the seismic data, e.g. ground roll noise.  For the correlated noise,  we generated a noise with a $1/|f|^\alpha$ spectral characteristic over its entire frequency range. Most common values for $\alpha$ are $\alpha = 1$ (pink noise) and $\alpha = 2$ (brown or brownian noise). Brownian noise is considered to be the worst correlated noise in seismic applications. Note that $\alpha=0$ refers to white noise. The autocorrelation and power spectrum density of various types of noises are shown in Fig. \ref{fig:specs}.

\begin{figure}
\centering
  \includegraphics[width=9cm]{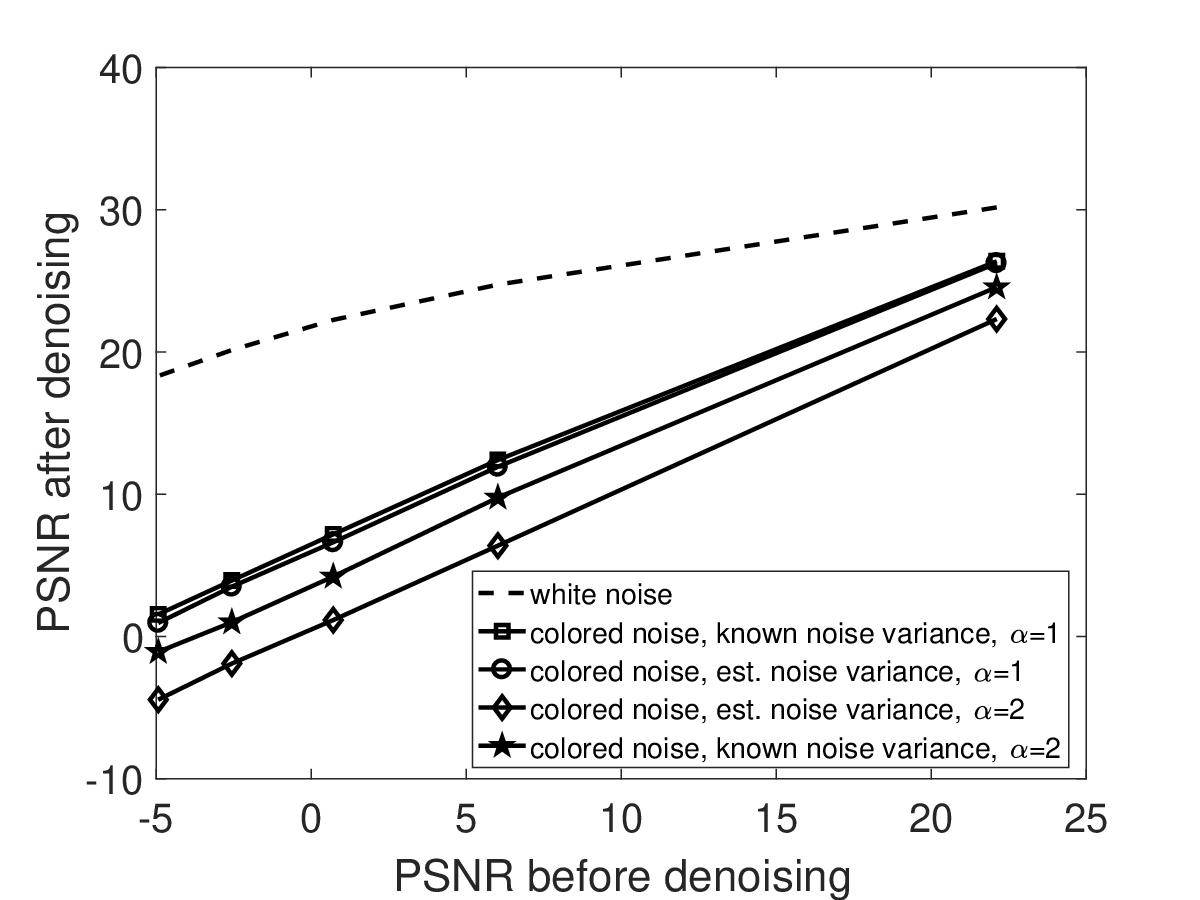}\\
  \caption{PSNR improvement for seismic images for various noises without whitening filter.}
  \label{fig:psnr_color}
\end{figure}

Comparison of the curvelet-based denoising under the correlated (color) and uncorrelated (white) noise reveals that the curvelet transform does not perform equally well in the later case (Fig.\ref{fig:psnr_color}). Hence, there is still room for the performance improvement. One reason of sub-optimal performance is that the noise samples are correlated and the other reason is the threshold which assumes white noise. Furthermore, for the case of correlated noise the noise level in each curvelet is different and, hence, defining a global threshold does not give best performance. One complex and difficult way is to identify the curvelets that contain  noise and define a curvelet dependent threshold based on the amount of noise in each curvelet. However, here we propose to include a pre-whitening filter the makes the noise to distribute equally among all the curvelets and  apply the noise estimation method describe previously to get the similar performance as uncorrelated noise. 
\subsection{Whitening filter}
\begin{figure}
\centering
  \includegraphics[width=9.1cm]{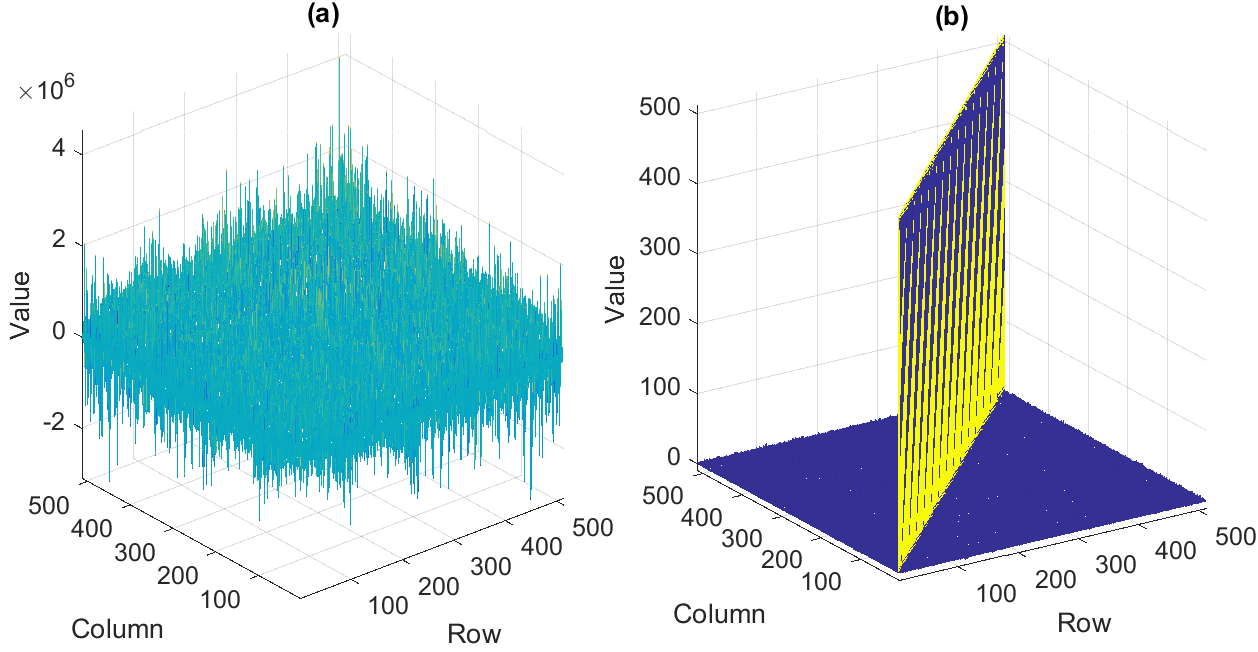}
  \caption{Covariance matrix of the noisy seismic image (a) before pre-whitening, (b) after pre-whitening.}
  \label{fig:auto}
\end{figure}

\begin{figure}
\hspace{-1cm}
  \includegraphics[width=18cm]{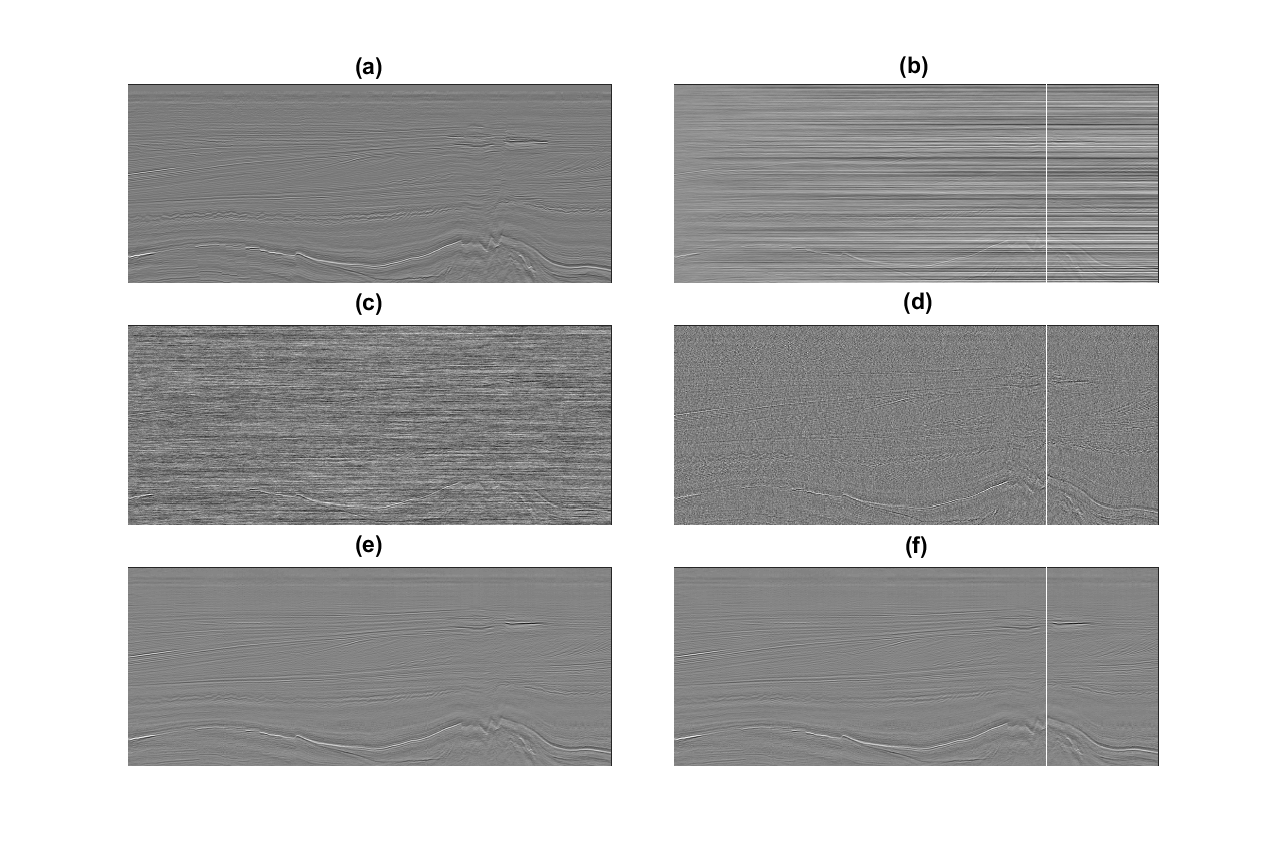}
  \caption{(a) Noise-less seismic image, (b) Noisy seismic image, $\alpha=2$, PSNR before denoising = $5$ dB. (c) Denoised image using curvelet without pre-whitening, (d) pre-whitened image, (e) denoised pre-whitened image together with whitening inverse ($\alpha=2$), (f) denoised pre-whitened image together with whitening inverse ($\alpha=1$).}
  \label{fig:prewhite}
\end{figure}

\begin{figure}
\centering
  \includegraphics[width=9cm]{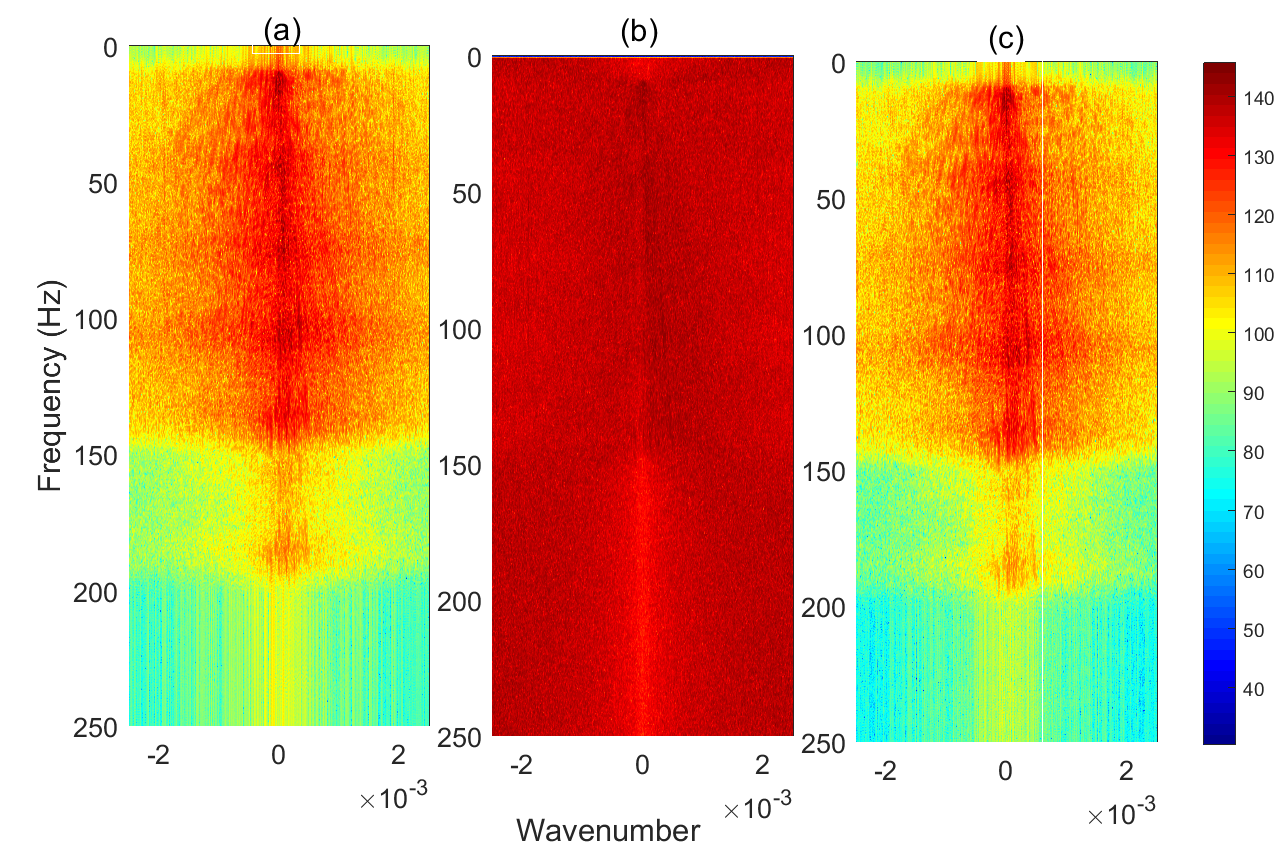}
  \caption{$f-k$ magnitude spectra: (a) Noise-less seismic data, (b) Noisy seismic data, $\alpha=2$, PSNR before denoising = $5$ dB. (c) Denoised seismic data.}
  \label{fig:fk}
\end{figure}

\begin{figure}
\centering
  \includegraphics[width=9cm]{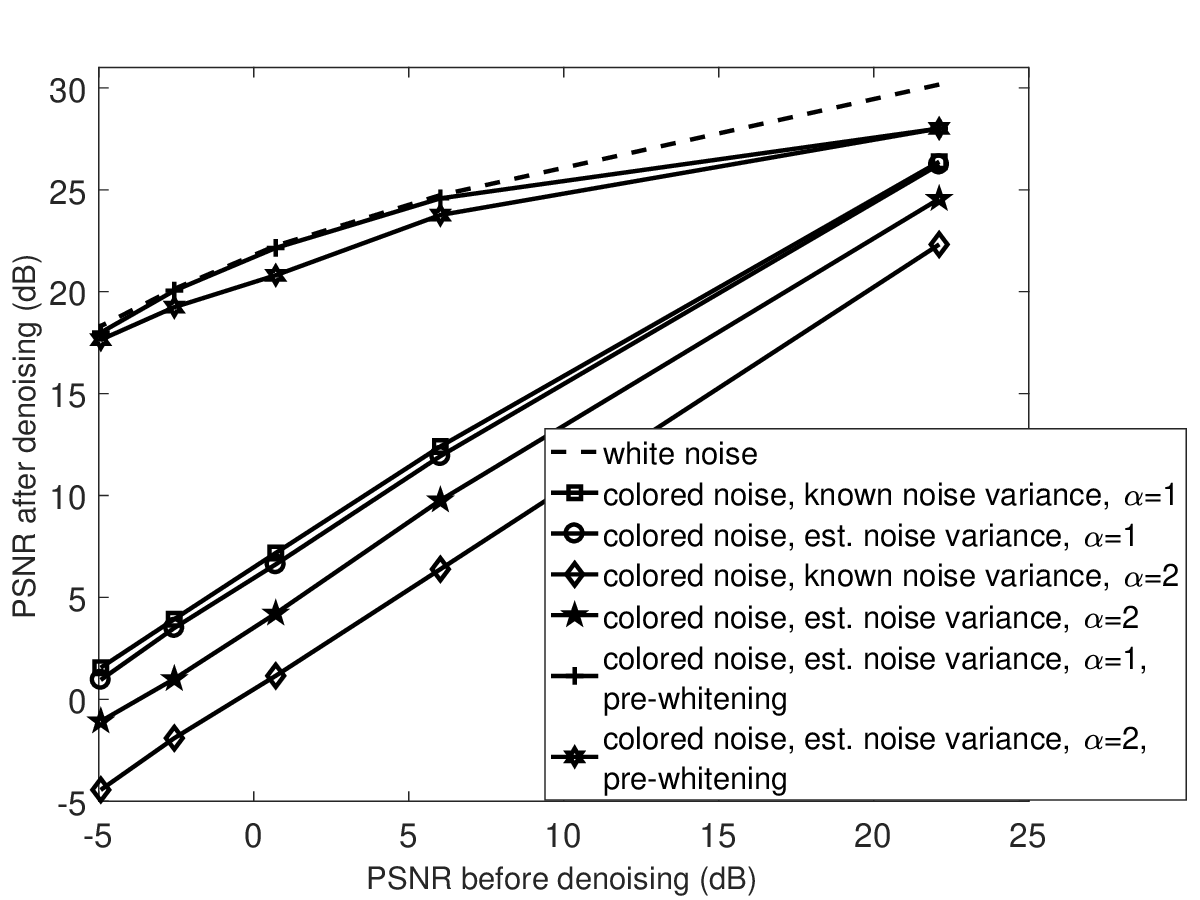}\\
  \caption{PSNR improvement for  seismic images with  various noises with pre-whitening.}
  \label{fig:psnr_color_white}
\end{figure}

To further improve the performance of curvelet-based denoising in correlated noise case, a pre-whitening filter (explained previously) is used. The pre-whitening filter makes the image uncorrelated and hence, application of curvelet transform based denoising gives better results  (close to white noise case). However, using pre-whitening on the noisy image some seismic signal information might be lost. For this purpose, after the denoising step using curvelet transform, whitening inverse filter is applied to get the final denoised image.
The noise estimation is performed as before using first $20 \times 20$ patch from the pre-whitened seismic image.
Figure \ref{fig:auto} shows the autocorrelation matrix of the image before and after whitening using ZCA. It can be seen from the figure that the covariance matrix after denoising is diagonalized.
The noisy and denoised (with and without whitening filter) is shown in Fig. \ref{fig:prewhite}. The performance difference of curvelet-based denoising with and without whitening can be clearly seen form the figure. From the Fig. \ref{fig:fk}, it can be seen that the $f-k$ magnitude spectra of the denoised seismic image closely matched with the $f-k$ spectra of noise-less seismic image.
The PSNR plot for the various noise cases and whitening method is depicted in Fig. \ref{fig:psnr_color_white}. The whitening method achieves close performance to that of the white noise case. The explicit PSNR values before and after denoising are shown in Table \ref{table:psnr}. Note here that the noise is estimated from the seismic image as before, which confirms that the noise estimation method gives best performance for incoherent noise case (either correlated or uncorrelated). In order to show the  performance superiority of the proposed method, the  technique is compared with the wavelet decomposition and empirical mode decomposition. For the wavelet decomposition based denoising technique, we  use ``wden'' function in the wavelet toolbox of Matlab \cite{Misiti1996}. Moreover, we use the principle of Stein's Unbiased Risk for soft thresholding (for details see  wavelet toolbox in Matlab2017 and references  \cite{Mousavi2016} and \cite{Stein1981} ). The comaprison is shown in Table \ref{tensor_comp}.
\subsection{Field data set with ground roll noise}
\begin{figure*}
\centering
  \includegraphics[width=15cm]{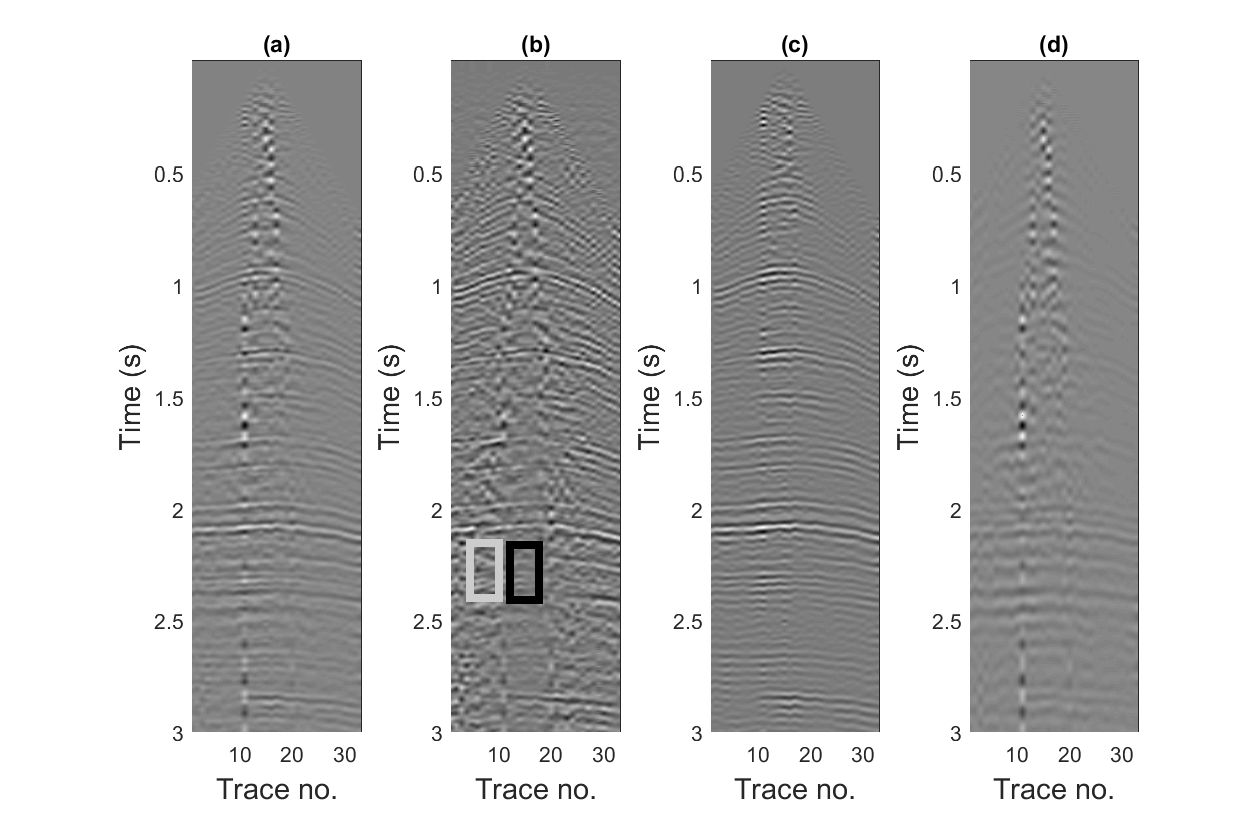}\\
  \caption{Seismic field data (shot gather number $8$). (a) before denoising, (b) after whitening, (c) after denoising and applying whitening inverse, (d) difference image of (a) and (c).}
  \label{fig:real}
\end{figure*}

To test the proposed method on field data with real noise (ground roll in this case), we used pre-stacked 2D land line from east Texas, USA. The ground roll noise have low apparent velocity, low frequency and high amplitude. For field data set, same procedure of denoising as before is used. First  data is whitened using whitening filtering. Second, the curvelet-based denoising is used, i.e., data is transformed to the curvelet-domain and threholding is used for noise attenuation. Third, denoised pre-whitened data obtained using inverse curvelet-transform. Finally, data is recovered using whitening inverse filter. Here, taking the  first $20 \times 20$ patch for noise estimation is not a good choice as in this patch there is no coherent  noise. The reason being that the noise is generated by the source (i.e., it is not a background noise like incoherent noise). For noise estimation in case of coherent noise, we proceed as follows: Considering the pattern of the ground roll noise seismic image, the patch at a furthest location directly under the shot location is assumed to be the signal only and the patch beside this is assumed to be contain the signal and noise. From these two patches the noise can be estimated using substraction. After substraction the variance (or standard deviation) of noise can be found for thresholding. The field data set, whitened data, denoised data and the difference between denoised and field data set is shown in Fig. \ref{fig:real}. The patches are shown in the Fig. \ref{fig:real}b by windows (signal-only part by black window, $S$  and noise plus signal part by grey window, $S_n$). Using the windows, the variance is calculated as
\begin{align}\label{sigma_c}
\sigma^2=E[(S_n(:)-S(:))^2]
\end{align}
where, $E[.]$ represents the statistical expectation, and $S_n(:)$ and $S(:)$ represents the elements of the matrix $S_n$ and $S$, respectively, arranged in a vector. Note that, in (\ref{sigma_c}) the independence between the signal and the noise cannot be assumed, since noise is a function of the source signal. 

In summary, whitening filter provides a threefold advantage. \begin{itemize}
\item First, noise estimation is carried out easily from the seismic image for coherent and incoherent cases. This is a  simple and straightforward method without much human intervention when compared to estimating the noise in the curvelets  domain by identifying the  curvelets corresponding to noise and  excluding them for reconstructing of the seismic signal  as done in \cite{Oliveira2012}. 
\item Second,  the performance of curvelet based denoising is improved for correlated and coherent noise cases (after whitening noise become random) and get close to the best denoising performance under the random noise case. It is known that denoising methods performs better in random noise case.
\item Third, the conventional hard thresholding method that assume white noise becomes valid. 
\end{itemize}
\section{Discussion and Conclusion}
Typical seismic data exhibit low SNR and correlated noise. Removing noise will drastically improve signal detection, seismogram composition
studies, and source discrimination for small local/regional seismic sources.  In this
work, we proposed a method based on the  curvelet transform. The curvelet transform is a higher dimensional
generalization of the wavelet transform, which is designed to represent images at different scales and different angles. It
basically overcomes the problem of missing directional selectivity of the wavelet transforms in images.
Comparison of the curvelet transform denoising performance for general purpose images and seismic
images shows that is better for seismic images. The reason being that the seismic data are more localized in time, frequency and phase which better match the properties of curvelet transform. The noise variance needed for the threshold was estimated from the image patches themselves. For incoherent correlated (pink and brown noise) and coherent noise cases, a   pre-whitening filter is introduced to enhanced the performance of curvelet denoising under colored noise. This makes the curvelet performs as good as in the case of white noise. Extensive testing of the proposed method on seismic images with incoherent (uncorrelated and correlated) and coherent  noise shows very promising performance.


\begin{table}[]\footnotesize
\centering
\caption{PSNR before and after denoising for various types of noises}
\label{table:psnr}
\begin{tabular}{|c|c|c|c|c|c|c|}
\hline
\multirow{2}{*}{\begin{tabular}[c]{@{}c@{}}PSNR before \\ denoising (dB)\end{tabular}} & \multicolumn{6}{c|}{PSNR after denoising (dB)} \\ \cline{2-7} 
 & \multicolumn{3}{c|}{\begin{tabular}[c]{@{}c@{}}without \\ pre-whitening\end{tabular}} & \multicolumn{3}{c|}{\begin{tabular}[c]{@{}c@{}}with \\ pre-whitening\end{tabular}} \\ \hline
 & $\alpha =0$ & $\alpha =1$ & $\alpha =2$ & $\alpha =0$ & $\alpha =1$ & $\alpha =2$ \\ \hline
-4.93 & 18.32 & 0.956 & -4.93 & 18.32 & 18.00 & 17.64 \\ \hline
-2.55 & 20.14 & 3.50 & -1.90 & 20.14 & 20.05 & 19.24 \\ \hline
0.71 & 22.26 & 6.62 & 1.15 & 22.26 & 22.15 & 20.81 \\ \hline
6.02 & 24.74 & 11.93 & 6.38 & 24.74 & 24.58 & 23.76 \\ \hline
22.11 & 30.16 & 26.24 & 22.32 & 30.16 & 28.01 & 28.01 \\ \hline
\end{tabular}
\end{table}
\begin{table}[t]
\footnotesize \centering
{
\caption{Comparison of  the proposed  method with other denoising methods.}
\label{tensor_comp}
{%
\begin{tabular}{|c|c|c}
\hline
Method                         & PSNR (dB)              \\ \hline
Noisy data set              & 0.710      \\ \hline
Wavelet decomposition        & 11.366 \\ \hline
Empirical Mode Decomposition  &  10.524 \\ \hline
Proposed method     &   22.26   \\ 
 \hline
\end{tabular}%
}}
\end{table}

\bibliographystyle{ieeetr}
\bibliography{library_fixed}
\end{document}